\documentclass[12pt]{iopart}  
\usepackage{graphicx}
\usepackage{harvard}
\usepackage[dvips]{color}  
\usepackage{times}
\begin{document}      
\jl{2}  
   
\title[Multiple scattering approach to low-energy electron collisions with
the water dimer] {Multiple scattering approach to elastic low-energy
electron collisions with the water dimer}
   
\author{D. Bouchiha,  L.~G.  Caron$^\dagger$, J.~D. Gorfinkiel$^\ast$ and L. 
Sanche}
   
\address{
Groupe en Science des Radiations, D\'epartement de M\'edecine Nucl\'eaire et 
Radiobiologie, Facult\'e de
M\'edecine, Universit\'e de Sherbrooke, Sherbrooke, Qc, J1H 5N4 Canada\\

$^\ast$Department of Physics and Astronomy, The Open University, Walton Hall, MK7 6AA Milton Keynes, UK\\

$^\dagger$D\'epartement de Physique et Regroupement Qu\'eb\'ecois sur
les Mat\'eriaux de Pointe, Universit\'e de Sherbrooke, Sherbrooke, Qc, J1K 2R1 
Canada}
   
\begin{abstract}   
Multiple scattering theory is applied to low-energy electron
collisions with a complex target formed of two molecular
scatterers. The total T-matrix is expressed in terms of the T-matrix
for each isolated molecule. We apply the approach to elastic
electron-(H$_2$O)$_2$ collisions. Following the method developed in
our previous work on crystalline ice \cite{KKR}, we impose a cut-off
on the dipole outside the R-matrix sphere and an energy dependent
cut-off on the angular momentum components of the monomer T-matrix. An
R-matrix calculation of electron-dimer collisions is performed in
order to evaluate the accuracy of the multiple scattering
approach. The agreement between the two calculations is very good.
\end{abstract}   

\section{Introduction}

In biological systems, water is present in a wide range of
environments. As a highly polar molecule, water tends to trap
low-energy electrons (LEEs) and is expected to play an important role
in LEE-induced processes \cite{Garrett2005,ptasinska2007}. Electron
collisions with water vapour being the simplest to study, extensive
experimental
\cite{Lozier1930,Schulz1960,Compton1967,Sanche1972,Belic81,Curtis92,Fedor2006,Rawat2007}
and theoretical \cite{Gil1994,Morgan1998,Gorfinkiel2002,Haxton2007b}
work has been carried out \citeaffixed{waterev}{for a comprehensive
review of work on electron scattering from gas phase water up to 2004
see}. In the more complicated condensed phase systems, experimental
studies of electron scattering from amorphous solid water, porous
amorphous solid water and crystalline ice have been reported
\cite{Rowntree1991,Simpson1997,Michaud2003,Orlando2005}. In water
clusters, formation of (H$_2$O)$_n^-$ by attachment of slow electrons
has also been investigated
\cite{Knapp1986,Knapp1987,Weber1999,Barnett1989,Lee1991} in order to
elucidate the mechanism of electron solvation.  To our knowledge, no
theoretical investigation of electron scattering with clusters or
condensed water have been reported; available methods are mainly
developed for electron collisions with gas-phase molecules. It is
therefore highly desirable to develop a new theoretical approach for
condensed-matter and cluster environments especially since many
electron-driven processes occur in these media.
 
The main idea behind the multiple scattering (MS) method is to
separate the potential of a complex target into non-overlapping
regions with each region taken as a single scatterer. The impinging
wave on each scatterer is composed of the incident plane wave and the
wave scattered from the other scattering centres. The advantage of
such an approach is that one can determine the cross section for large
systems by combining information from its subunits. This enables the
study of larger targets (macromolecules, molecular clusters, etc.) 
than is possible using the methods currently available.

The MS approach to LEE scattering was previously proposed by two of
the authors (L. Caron and L. Sanche) \cite{PRL,PRA,PRA2} and its use
in conjunction with R-matrix calculations was recently validated
\cite{KKR} (hereafter referred to as I). The proposed approach combines
scattering gas-phase data, obtained from accurate {\it ab initio}
R-matrix calculations, to derive scattering information for a
condensed-matter situation. The partitioning of the space in the
R-matrix method makes it ideal for a multiple scattering approach of
the "muffin-tin" type. The method is forcibly less rigorous than the
ones available for small gas-phase targets, but it is fast and, for
the systems tested so far, yields very good results.

In I, we have used the MS theory in the condensed phase, using the
equivalent KKR (Korringa, Kohn, and Rostoker) approach, in combination
with gas-phase R-matrix data in the form of T-matrices. We were able
to derive the band structure of a molecular crystal (ice). The present
work constitutes a first attempt to determine elastic cross section
(CS) for LEE scattering from a molecular cluster using the MS
approach. The water dimer is an excellent choice for such an attempt,
primarily because, making use of our previous experience in
electron-H$_2$O collisions \cite{Gorfinkiel2002}, we were able to
perform accurate R-matrix calculations of electron-dimer collisions and test the MS technique.
In addition, water is present in a large variety of environments and
information on LEE interactions with it is highly relevant. For
example, the characteristics of the water in some of the solvation
layers surrounding the double-helix structure of DNA are not that of
bulk water \cite{Becker1997}, so may be explained in terms of
clustered H$_2$O.

The main question we address in this paper is whether a
multiple scattering approach is a good alternative to the
computationally heavy {\em ab initio} methods for large to very large
targets. For this purpose, we investigate how to apply the cut-offs
found in I in order to perform a MS calculation of LEE collision with
the water dimer.

\section{Theory}

In this section we present a multiple scattering method to calculate
the total T-matrix, $\mathbf{T}^{tot}$, describing LEE collisions from a target
consisting of two identical scatterers from their individual T-matrices $\mathbf{T}$. 
The monomers, labelled with the index $n$, are located
at $\vec{R}_{n}=\vec{R}_{\pm}=\pm \vec{R}$.

The asymptotic wave function far away from the centre of mass of the
two molecules can be expressed as follows:

\begin{equation}
\psi (\vec{r})=\sum_{LL^{\prime }}Y_{L^{\prime }}(\hat{r})\left[
j_{l^{\prime }}(\kappa r)\delta _{LL^{\prime }}+\frac{1}{2}h_{l^{\prime
}}^{(+)}(\kappa r)T_{L^{\prime }L}^{tot}\right] f_{L}^{0}~.  \label{eq_0}
\end{equation}
where $r$ is the relative coordinate of the scattered electron and the
centre of mass of the target. The $Y_{L}$ are spherical harmonics with
$L=(l,m)$, $j_l$ denotes the spherical Bessel functions and
$h_{l}^{(+)}$ the spherical Hankel functions of the first
kind. $\kappa=\sqrt E$ with $E$ the energy of the incoming electron
and $f_{L}^{0}$ is an amplitude factor for the incident plane
wave. The first term on the right side of (\ref{eq_0}) corresponds to
the total incident wave and the second to the total scattered wave.

We need to re-expand the incident wave in (\ref{eq_0}) around the
centre of mass of each scatterer $n$. We use the general re-expansion
formula from \citeasnoun{DillDehmer} which, for the spherical
harmonics defined by Messiah \cite{Messiah}, is

\begin{equation}
\fl Y_{L^{\prime }}(\hat{r})h_{l^{\prime }}^{(\pm )}(\kappa
r) =\sum_{L_{1},L_{2}}i^{l_{1}+l_{2}-l^{\prime }}(-1)^{m^{\prime
}}F_{m_{1,}m_{2,}-m^{\prime }}^{l_{1},l_{2},l^{\prime }}Y_{L_{1}}\left( \hat{%
r}_{n}\right) Y_{L_{2}}\left( \hat{R}_{n}\right)
h_{l_{1}}^{(\pm )}(\kappa
r_{n})j_{l_{2}}\left( \kappa R_{n}\right) ~  \label{eq_1}
\end{equation}%
where $\vec{r}_{n}=\vec{r}-\vec{R}_{n}$ and%
\begin{equation}
\fl F_{m_{1},m_{2},m_{3}}^{l_{1},l_{2},l_{3}} =\left[ 4\pi
(2l_{1}+1)(2l_{2}+1)(2l_{3}+1)\right] ^{\frac{1}{2}}\left( 
\begin{array}{ccc}
l_{1} & l_{2} & l_{3} \\ 
0 & 0 & 0%
\end{array}%
\right) \left( 
\begin{array}{ccc}
l_{1} & l_{2} & l_{3} \\ 
m_{1} & m_{2} & m_{3}%
\end{array}%
\right).
\end{equation}%
$\left( 
\begin{array}{ccc}
l_{1} & l_{2} & l_{3} \\ 
m_{1} & m_{2} & m_{3}%
\end{array}%
\right) $ is the Wigner 3-j symbol. 

\bigskip

It follows that the incident wave on each scatterer is:

\begin{equation}
\psi_{i} (\vec{r}_{n})=\sum_{L^{\prime }}Y_{L^{\prime }}(\hat{r}%
_{n})j_{l^{\prime }}(\kappa r_{n})g_{nL^{\prime }}^{0}
\end{equation}%
where 
\begin{equation}
g_{nL_{1}}^{0}=\sum_{L}M1_{L_{1}L}^{n}f_{L}^{0}
\end{equation}%
or, in matrix form
\begin{equation}
\mathbf{g^{0}}=\mathbf{M1\cdot f^{0}}
\label{eq_M1}
\end{equation}%
with
\begin{equation}
M1_{L_{1}L}^{n}=%
\sum_{L_{2}}i^{l_{1}+l_{2}-l}(-1)^{m}F_{m_{1,}m_{2,}-m}^{l_{1},l_{2},l}Y_{L_{2}}\left( 
\hat{R}_{n}\right) j_{l_{2}}\left( \kappa R_{n}\right) ~.
\end{equation}

\bigskip

We can now incorporate the multiple scattering between the two
scatterers to obtain a global expression for the total impinging 
wave on each molecule. This has the form
\begin{equation}
\psi_{i} (\vec{r}_{n})=\sum_{L}Y_{L}(\hat{r}%
_{n})j_{l}(\kappa r_{n})g_{nL} .
\label{eq_g1}
\end{equation}%
One can identify our $g_{nL}$ with the expression $4 \pi i^{l} e^i
\vec{k}\cdot \vec{R}_{n} B_{\vec{k} L} ^{(n)}$ in the analysis of 
\citeasnoun{PRA}. By comparing Eq.~(\ref{eq_g1}) to the total 
impinging part of their Eq.~(1), {\it i.e.} the first term on the
right-hand side, one then gets from their Eq.~(2)

\begin{eqnarray}
g_{nL} &=&g_{nL}^{0}+ \frac{1}{2}\sum_{L_{1},L_{2},L_{2}^{\prime
}}i^{l_{1}+l-l_{2}^{\prime }}T_{L_{2}^{\prime }L_{2}}^{n^{\prime
}}g_{n^{\prime }L_{2}}    
(-1)^{m_{2}^{\prime }}F_{m_{1,}m_{,}-m_{2}^{\prime
}}^{l_{1},l,l_{2}^{\prime }}Y_{L_{1}}\left(\hat{R}_{nn^{\prime}}\right)
h_{l_{1}}^{+}\left( \kappa R_{nn^{\prime }}\right)    \nonumber \\
&=&g_{nL}^{0}+\sum_{L_{2}L_{2}^{\prime }}\chi _{LL_{2}^{\prime
}}^{nn^{\prime }}T_{L_{2}^{\prime }L_{2}}^{n^{\prime }}g_{n^{\prime }L_{2}}
\label{eq_g}
\end{eqnarray}
with $\vec{R}_{nn^{\prime}} = \vec{R}_n - \vec{R}_{n^{\prime}}$,
$n^{\prime }=-n$.
\begin{equation}
\chi _{LL_{2}^{\prime }}^{nn^{\prime
}}=\frac{1}{2} \sum_{L_{1}}i^{l_{1}+l-l_{2}^{\prime }}(-1)^{m_{2}^{\prime
}}F_{m_{1,}m_{,}-m_{2}^{\prime }}^{l_{1},l,l_{2}^{\prime }}Y_{L_{1}}\left( 
\hat{R}_{nn^{\prime }}\right) h_{l_{1}}^{+}\left( \kappa R_{nn^{\prime}}\right)
\end{equation}
expresses how a scatterer influences its neighbour. Note that for
$n=n^{\prime}$ $\chi _{LL^{\prime }}^{nn}=0$. The second term on the
right side of (\ref{eq_g}) is the total scattered wave from the second
scatterer $n^{\prime }$.  In matrix form, (\ref{eq_g}) can be written
as
\begin{equation}
\mathbf{g}=\left( \mathbf{I}-\mathbf{\chi T}\right) ^{-1}\cdot \mathbf{g^{0}}.~  
\label{eq_chi}
\end{equation}
Now, we have to combine the amplitudes scattered from each scatterer
\begin{equation}
\psi _{s}(\vec{r})=\sum_{n}\psi _{s}(\vec{r}_{n})=\sum_{n}\sum_{LL^{\prime
}}Y_{L^{\prime }}(\hat{r}_{n})\frac{1}{2}h_{l^{\prime }}^{(+)}(\kappa
r_{n})T_{L^{\prime }L}^{n}g_{nL}~.
\end{equation}%
around the common origin (centre of mass) using
\begin{equation}
\fl Y_{L^{\prime }}(\hat{r}_{n})h_{l ^{\prime }}^{(\pm )}(\kappa
r_{n})=\sum_{L_{1},L_{2}}i^{l_{1}+l_{2}-l^{\prime }}(-1)^{m^{\prime
}}F_{m_{1,}m_{2,}-m^{\prime }}^{l_{1},l_{2},l^{\prime }}Y_{L_{1}}\left( \hat{%
r}\right) Y_{L_{2}}\left( -\hat{R}_{n}\right) h_{l_{1}}^{(\pm )}(\kappa
r)j_{l_{2}}\left( \kappa R_{n}\right).  \label{eq_2}
\end{equation}%
One gets%
\begin{equation}
\psi _{s}(\vec{r})=\sum_{L_{1}}Y_{L_{1}}(\hat{r})\frac{1}{2}%
h_{l_{1}}^{(+)}(\kappa r)\sum_{n,L,L^{\prime }}M2_{L_{1}L^{\prime
}}^{~~~n}T_{L^{\prime }L}^{n}g_{nL}~  \label{eq_M2}
\end{equation}%
where%
\begin{equation}
M2_{L_{1}L^{\prime }}^{~~~n}=\sum_{L^{\prime
},L_{2}}i^{l_{1}+l_{2}-l^{\prime }}(-1)^{m^{\prime
}}F_{m_{1,}m_{2,}-m^{\prime }}^{l_{1},l_{2},l^{\prime }}Y_{L_{2}}\left( -%
\hat{R}_{n}\right) j_{l_{2}}\left( \kappa R_{n}\right) ~.
\end{equation}%
Combining (\ref{eq_M2}), (\ref{eq_M1}), (\ref{eq_chi}) and referring
to the general expression (\ref{eq_0}), we have
\begin{equation}
\mathbf{T}^{tot}=\mathbf{M2\cdot T\cdot} (\mathbf{I}-\mathbf{\chi T})^{-1}\cdot 
\mathbf{M1}~.
\label{eq:final}
\end{equation}

We thus obtain the T-matrix for two identical scatterers from their
individual T-matrices.  {\bf M1} and {\bf M2} account for the
transformation from the monomer to the dimer's centre of mass and
$(\mathbf{I}-\mathbf{\chi T})^{-1}$ expresses the multiple scattering
between the two scatterers.

\section{Characteristics of the calculations}

No experimental or theoretical data on elastic LEE collisions with the
water dimer are available. Therefore, in order to test the quality of
our MS results, we have performed an {\em ab initio} scattering
calculation on the gas-phase dimer using the R-matrix method and the UK
polyatomic R-matrix suite \cite{MTG98}.

\begin{figure}
\begin{center}
{\resizebox{130mm}{!}{ \includegraphics*{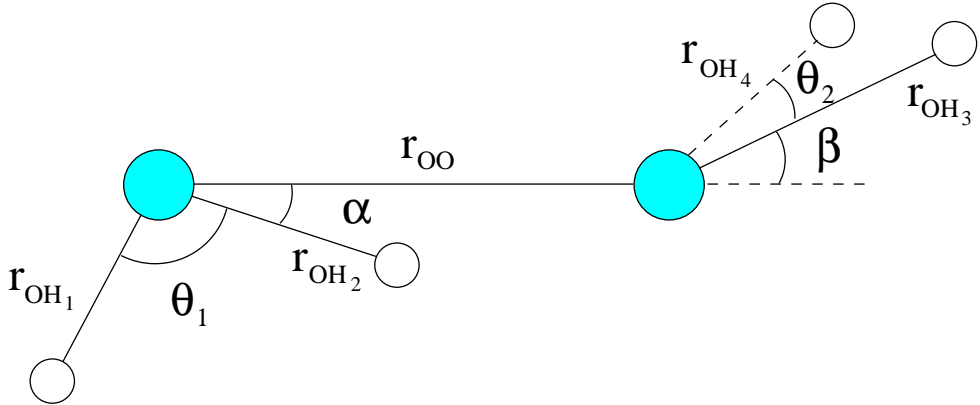}}}
\caption[]{Ground state equilibrium geometry of (H$_2$O)$_2$. The parameters' values 
are listed in Table~\ref{tab:geometry}.
\label{fig:geometry}}
\end{center}
\end{figure}

The gas-phase water dimer in its ground state equilibrium geometry has
$C_s$ symmetry (Figure~\ref{fig:geometry}). The geometry parameters
from \citeasnoun{PKK2001} were used in the R-matrix calculation. The
geometry of the water monomer is very slightly changed upon dimer
formation (see Table~\ref{tab:geometry}). In the MS calculation, the
dimer was formed by putting together two water molecules separated by
$d \sim 5.5$~bohr along the Z-axis. This means that there are very
minor differences between the geometries used in the R-matrix and MS
calculations. The most notable is the hydrogen bond r$_{OH}$ which is
at most $\sim 0.013$~bohr longer in the dimer than in the monomers we
use to build the MS results.

\subsection{R-matrix calculation for the dimer}

A detailed description of the R-matrix method as applied to polyatomic
molecules within the fixed-nuclei (FN) approximation can be found in
\citeasnoun{MGT97} and \citeasnoun{MTG98}. The method is based on
splitting coordinate space into two regions separated by a sphere,
henceforth called R-sphere, centred on the centre of mass of the
molecule. The radius of the R-sphere is chosen in such way that the
electronic density of the target is negligible outside it. As a
consequence, exchange and correlation effects can be neglected in this
outer region and a long-range multipole expansion used to represent
the electron-target interaction. Inside the R-sphere, both effects are
significant and are therefore taken into account using rigorous
quantum chemistry methods. The wavefunction for the target~+~electron
system is then expressed in terms of a close-coupling expansion.

The calculation was performed following previous work on electron
scattering from isolated H$_2$O \cite{Gorfinkiel2002}. We used the
equilibrium geometry from \citeasnoun{PKK2001} and the same basis set (including  
the diffuse functions) employed for the monomer work by \citeasnoun{Gorfinkiel2002}. 
Since in this work we are concerned with the elastic scattering, only the ground electronic state was
considered and included in the close-coupling expansion. We therefore
produced natural orbitals exclusively from this state and then used
them in a CASCI (complete active space configuration interaction)
calculation in which 8 electrons are frozen; this generated around
7000 configurations. With this model, we obtained a good value for the
ground state energy: -152.18~hartree, compared to -152.67~hartree from
the most accurate calculation. We also obtained excellent agreement
with the experimental dipole moment: 1.065~a.u. in our calculations
compared to 1.041~a.u..

We used an R-sphere radius of $a$=13~bohr.  In order to confirm that
all the electronic density was contained inside the sphere, tests were
performed for radii of $a$=14 and $a$=15~bohr. Only small differences
at very low energies were found between the cross sections calculated
with the three radii \cite{jimena}. The continuum orbitals describing
the scattered electron were expanded in a basis of GTOs with $l \leq
4$ centred on the centre of mass of the water molecule.

\subsection{Multiple scattering calculation}

In I, we showed that gas-phase R-matrix data can be efficiently used
in conjunction with MS theory in a non-gaseous environment by deriving
the band structure of a molecular crystal (ice). The fundamental
lesson learnt from this KKR study is that for the multiple scattering
term, we need a trimmed T matrix $\mathbf{T_c}$, without dipole
contribution for radii $r > a_c$ and with an angular momentum
cut-off. Thus two different monomer T-matrices are needed:
$\mathbf{T_c}$ for the multiple scattering term $(\mathbf{I}-\chi
\mathbf{T_c)}^{-1}$ and a matrix $\mathbf{T_{dip}}$ for which no
cut-offs are applied. With this in mind Eq.~\ref{eq:final} can be
rewritten:
\begin{equation}
\mathbf{T}=\mathbf{M2}\cdot \mathbf{T_{dip}}\cdot (\mathbf{I}-\chi \mathbf{T_c)}^{-1}\cdot \mathbf{M1}~.
\label{eq:used}
\end{equation}

The first step in our multiple scattering calculation involves
generating the monomer T-matrices $\mathbf{T_c}$ and
$\mathbf{T_{dip}}$ required in equation~(\ref{eq:used}). The
description of LEE scattering by an isolated H$_2$O molecule is the
same than in I. Since a detailed description of the calculation was
given there, we will limit ourselves here to a brief summary. We have used the
R-matrix method within the fixed-nuclei approximation. For the target
description we used the DZP basis set \cite{Dun} for the Oxygen and a
TZ basis \cite{Dun71} for the Hydrogen.  A CASCI model was used with
only 2 frozen electrons and the calculation included only the ground
state. The scattering calculation was performed with an R-sphere
radius of $a = 6$~bohr. A GTO basis for the continuum with $l
\leq 4$ was generated for this radius using the program GTOBAS
\cite{FGM01} .  The obtained T-matrices were rotated from the R-matrix
coordinates to their corresponding position in the dimer. The
procedure to do so was explained previously (Equation~(24) in I) and
the reader is referred to that work for further details.

As mentioned above, two cut-offs are applied when generating
$\mathbf{T_c}$.  A first cut-off on the range of the molecular dipole
is achieved by removing the dipole field for $r > a_c$ . The cut-off
radius $a_c$ depends only on the dipole moment of the molecular
monomer and should therefore be the same for the ice and the dimer
calculations. In I, we have found that $a_c \sim 6.5$~bohr yields the
best value for the electron's effective mass in ice.  An analysis of
the electron-(isolated) water cross section obtained for different
values of $a_c$ supports this finding (the reader is referred to
Fig. 1 in I): for too small cut-off radii ($a_c < 6$), the high energy
cross section is unphysical whereas for higher radii ($a_c > 6$) the
cross section exhibits a dipole-driven behaviour (it increases rapidly
at low energy). Although this behaviour is physical and should be
expected, it is the aim of the cut-off to at least partially eliminate
it.  This can be understood as an attempt to ensure that
$\mathbf{T_c}$ involves only the contribution of the target potential
in the small region around the scatterer.  That leaves $a_c
\sim 6$~bohr as the most satisfactory radius.

The second cut-off, in the angular momentum components $l$ of the
scattering matrix, is critical and depends both on the electron
kinetic energy $E$ and the intermolecular distance $d$. The
restriction is related to the angular momentum energy barrier
$E(l,r)=l(l+1)/r^2$. The KKR calculation predicts that only angular
momentum values $l \leq l_{c}$ should be retained such that
$E(l_{c},d) < E_{e} < E(l_{c}+1,d)$ with $E(l,d)=l(l+1)/d^{2}$
Ry. That is, only electrons that scatter from one monomer with energy
larger than $E(l,d)$ can reach the other one and undergo multiple
scattering. In the dimer, the two H$_2$O molecules are separated by $d
\sim 5.5$~bohr (almost the same intermolecular distance than in
crystalline ice in I). We expect then to have the same cut-off
criteria than in the KKR calculation.

Once the T-matrices for the monomer are calculated, the dimer T-matrix
in equation~\ref{eq:used} is built. From it, we build the R-matrix at
$r=a_{dip}=9$~bohr using the approach given in Appendix A and
propagate it outwards \cite{Baluja1982} using \textit{the exact
dipolar field of the dimer}. This is necessary in order to incorporate
the effect of the dipole moment of the dimer in the MS description.
To choose $a_{dip}$ we propagate the R-matrix of the dimer (obtained
with the R-matrix calculation) inwards from the asymptotic region and
then calculate the CS from it. The low-energy part of this CS
decreases the more of the dipole field we remove (the further in we
go) and takes an unsuspected and incorrect upturn for a radial
distance of 9~bohr. We believe that for this distance the multipolar
field of the dimer is no longer describable by its total dipole
field. We finally generate the final T-matrix $\mathbf{T}^{tot}$ which
takes the dimer's dipole contribution into account. This T-matrix is a
function of energy and of the angular momentum cut-off $l_c$ used for
$\mathbf{T_c}$ in the multiple scattering term of
Eq.~\ref{eq:used}. The integral elastic cross section is then obtained
using the well known formula:

\begin{equation}
\sigma(E) = \frac{\pi}{k^2}  \sum_{l l'} |T^{tot}_{ll'} |^2
\end{equation}

\begin{figure}
\begin{center}
{\resizebox{130mm}{!}{ \includegraphics*{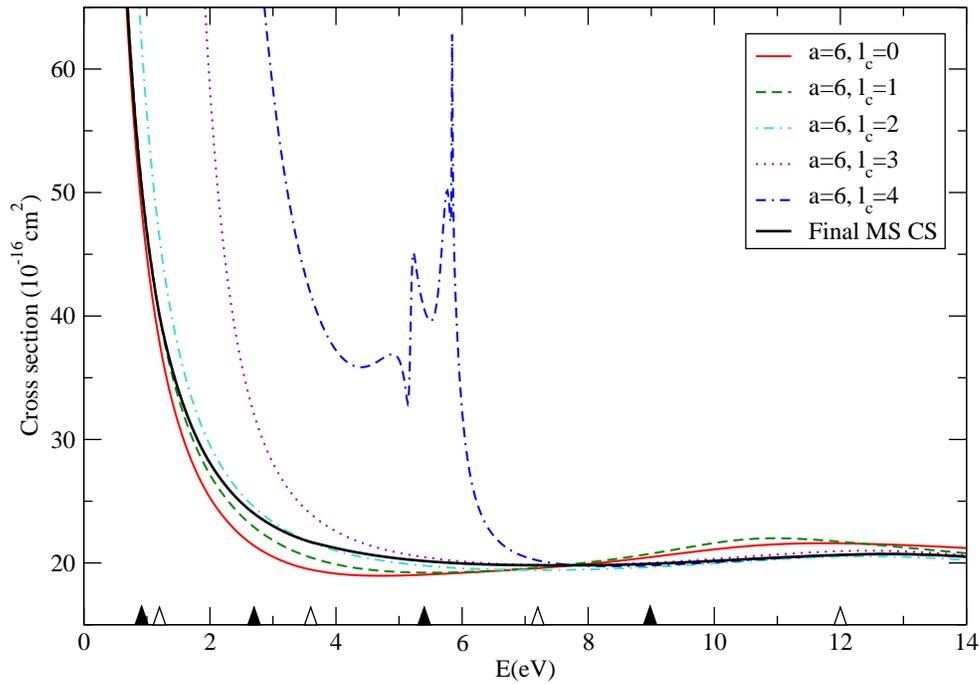}}}
\caption[]{Trial elastic electron-water dimer cross sections calculated 
using different angular momentum cut-offs $l_c$. The MS term in
Eq.~(\ref{eq:used}) introduces unphysical resonances and/or premature
divergences in the CS. The filled arrowheads point to the cut-off
energies $E(l_{c},d)$ while the empty arrowheads point to the shifted
regularised energies $E_{s}(l_{c},d)$ for $l_{c}=1,2,3,4$
respectively, from left to right.
\label{fig:test_l}}
\end{center}
\end{figure}

In Figure~\ref{fig:test_l}, we present the CS for different $l_c$
values. As mentioned earlier, the partial wave expansion must be
limited as the collision energy decreases.  For example, for electron
energies between $5.4 < E <9$ eV, the KKR calculation predicts
that we can only include $l \leq 3$ in the MS calculation. It can be
easily seen from Figure~\ref{fig:test_l}, that the CS is ill behaved in 
this region (below $\sim 7$ eV) when $l_c=4$. A closer look at the CS
for the other values of $l_c$ shows the presence of a critical energy
below which the CS does not exhibit the right behaviour. These
critical energies are at slightly lower energies than the value given by
$E(l_{c},d)$.

If we scrupulously restrict the calculation to integer values of $l_c$
for electron energies between $E(l_{c},d)$ and $E(l_{c}+1,d)$, the CS
will be jagged {\it i.e.} discontinuous at $E(l_{c},d)$. Smoothing is
needed. A two-point interpolation on the parameter $l$ between $l_{c}$
and $l_{c}+1$, where $l$ is the solution of $E=l(l+1)/d^{2}$, would
get rid of the discontinuities. But this would admix the CS at $l_{c}$ 
with any singular part of the CS at $l_{c}+1$. A
quick look at Figure~\ref{fig:test_l} indicates, for instance, that at
6 eV when $l_{c}=3$ the interpolation would spuriously bring in a
sizable part of the huge peak of the $l_{c}=4$ CS. The procedure can
be regularised by upward shifting $E(l_{c},d)$ using a parameter
$\gamma$ such that $\gamma E_{s}(l_{c},d)=l_{c}(l_{c}+1)/d^{2}$. A
value of $\gamma = 0.75$ would shift the cut-off energies to the new
values shown in Figure~\ref{fig:test_l}, beyond the threshold of the
singular behaviour of the CS for $l_{c}+1$. The interpolation procedure
can now be done safely with $l$ a solution of $\gamma
E_s=l(l+1)/d^{2}$. The CS is linearly interpolated between CS values
corresponding to the closest lower and larger integer values of $l$.
 
\section{Results}

\begin{figure}
\begin{center}
{\resizebox{130mm}{!}{ \includegraphics*{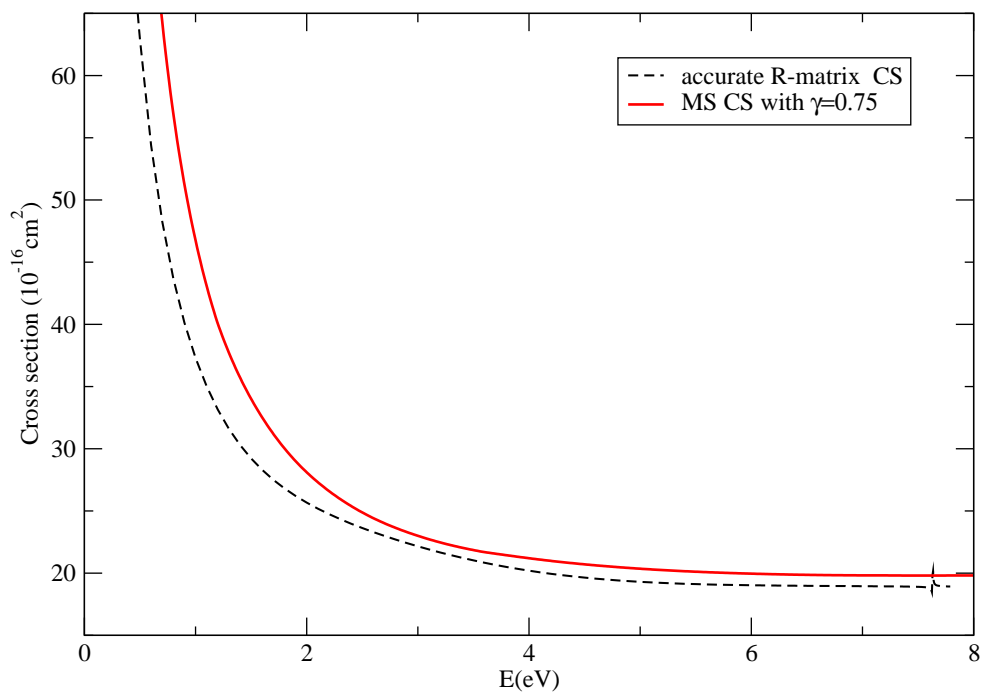}}}
\caption[]{Elastic electron-(H$_2$O)$_2$ cross sections. The MS cross section 
was obtained following the smoothing prescription explained in the
text. Notice that he first vertical excitation threshold is located
around 6.99~eV \cite{VvG05}.
\label{fig:Xs_final}}
\end{center}
\end{figure}

The elastic electron-(H$_2$O)$_2$ cross section versus impact energy is plotted in Figure~\ref{fig:Xs_final}. As a test of the accuracy of the MS calculation, we compare our results to the CS calculated with the R-matrix codes. The agreement between the two calculation is very good. The multiple scattering cross section is slightly higher than the R-matrix one but remains within 5\% of the the fully {\em ab initio} CS for energies larger than 2.5 eV. We believe these differences are well within the range of usual experimental errors and ab initio calculation uncertainties.  For targets with large dipole moments, it is customary to add to the R-matrix cross section a Born based correction \cite{born} to account for the partial waves not included in the {\it ab initio} calculation. This correction is the same for
the R-matrix and MS calculations and for this reason has not been included in the cross sections plotted in Figure~\ref{fig:Xs_final}.

Based on the results of I, the need for a ``trimmed'' T-matrix ($\mathbf{T_c}$) in the MS term of (\ref{eq:used}) is obvious. Figure~\ref{fig:test_l} illustrates the drastic impact of a larger angular momentum basis; including higher $l$ at low energies introduces resonances and/or premature divergences in the CS caused by the resonant nature of the MS $(\mathbf{I}-\chi \mathbf{T)}^{-1}$ term in Eq.~\ref{eq:used}. The cut-off in the long-range dipolar interaction is also very important. In I, we have shown that even though the value of the cut-off $a_c$ is not critical for band structure calculations, a too large or too small cut-off radius has a noticeable effect on the value of the effective mass (which was used for calibration purposes).  The two cut-offs are necessary and have to be applied only to the MS term. The use of a
full T-matrix $\mathbf{T_{dip}}$ in (\ref{eq:used}) is, however, essential to get the right CS behaviour at low energies and to properly account for the potential produced by the dipole of a H$_2$O molecule. The need for $\mathbf{T_{dip}}$ to include the dipole contribution arises from the fact that for $r \leq a_{dip}$, the scattering electron feels the individual dipoles of the isolated water molecules.

\begin{figure}
\begin{center}
{\resizebox{80mm}{!}{ \includegraphics*{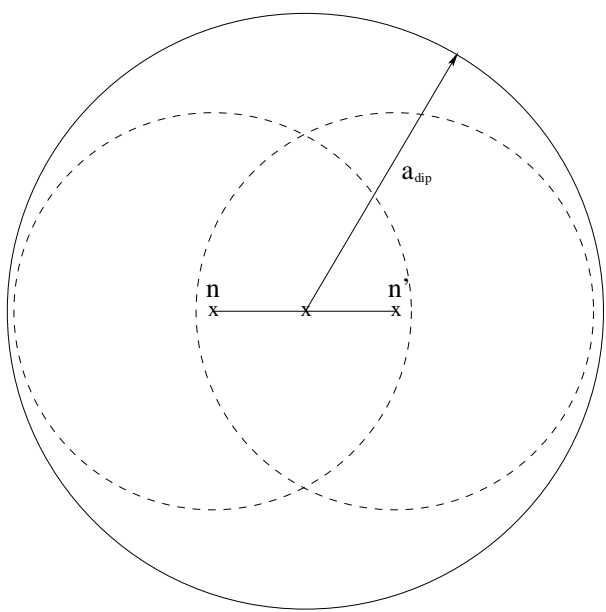}}}
\caption[]{Radii used in the MS calculations; solid line: $a_{dip}= 9$~bohr, for inclusion of the dimer's dipole effect. Dashed line: R-sphere
with radius $a=6$~bohr used in the R-matrix calculation for the
monomer. $n$ and $n^{\prime}$ define the positions of the centre of
mass of each water molecule.
\label{fig:r-spheres}}
\end{center}
\end{figure}

The radius $a_{dip}$ has to be chosen in a proper way. It must allow
the separation between the regions where the individual water dipoles
dominate the electron-molecule interaction ($r < a_{dip}$) and the
region where the total dipole of the dimer dominates ($r >
a_{dip}$). Based on a geometrical analysis
(Figure~\ref{fig:r-spheres}), one can see that a radius of 9~bohr is
an adequate choice.  In order to check the appropriateness of this
value, we have tried several R-sphere radii ranging from 6 to 13~bohr
and confirmed that 9~bohr yields the best cross section.

It is evident from Figure~\ref{fig:r-spheres}, that a choice of $a
\simeq a_c \simeq 6$~bohr in our calculations does not exactly
correspond to a non-overlapping potentials picture. For the spheres
not to overlap we would need a radius around $d/2 \simeq
2.7$~bohr. However, using such small radius does not provide a good
representation of the target. This was already clear in I, where we
established that small radii introduce a discontinuity at the boundary
of the sphere that is too important to lead to good results.

\section{Conclusions and Discussion}

We have developed a technique to derive the total T-matrix of two scatterers from their individual T-matrices as described by Eq.~(\ref{eq:final}). The technique first involves finding the T-matrix describing the scattering for an isolated molecular target. These are then combined using Eq.~\ref{eq:used}. Our main conclusion is that two different T-matrices, with different constraints, are needed to represent the electron-dimer interaction: two cut-offs have to be applied only to the MS term ($\mathbf{T_c}$ but not $\mathbf{T_{dip}}$). From the total T-matrix, the R-matrix at $a_{dip}$ is determined and then propagated to an asymptotic distance using the dipole moment of the cluster. This propagation is fundamental in order to incorporate the electron interaction with the true dimer dipole moment. One can argue that the need for $\mathbf{T_{dip}}$ and $a_{dip}$ is a surface effect not present in the infinite crystal situation in I. In fact, the dimer is mostly surface.
   
Our results show that a multiple scattering calculation can efficiently replace a \textit{standard ab initio} scattering calculation for this system. Very good agreement is found between the MS and the R-matrix cross sections. The MS method should allow us to calculate elastic cross sections for small water clusters that are nevertheless too big to be studied {\it ab initio}. The same value of $a_c$ should be employed and $a_{dip}$ could be determined using the same geometrical considerations. How large a cluster could one attempt to treat using this procedure? The cluster should have a large surface to volume ratio. The arguments presented in Appendix B indicate that the number of water molecules should be much less than 500. That leaves a lot of room for fair sized clusters.

For very large clusters, having a low surface to volume ratio, the
general picture would be as follows. The physical space should be
divided into three parts. In the innermost part, the bulk of the
cluster, MS theory using $\mathbf{T_c}$ in a uniform background
optical potential $U_{op}$, measuring the average polarisation
potential energy between the muffins, would be used as in I. Note that
the resonant MS $(\mathbf{I}-\chi \mathbf{T_c)}^{-1}$ term of
Eq.~(\ref{eq:used}), which has been freed of any spurious peaks
through the angular momentum cut-off algorithm, will yet show peaks
that correspond to damped electronic modes of the cluster. These are,
for instance, the modes that were calculated for crystalline ice in
I. Then there would be a narrow surface region which adapts the bulk
value of the optical potential to the vacuum level and in which
$\mathbf{T_c}$ would be used for the MS part and $\mathbf{T_{dip}}$
for the exit part. Finally, there would be an outer region in which
the R-matrix of the two inner regions can be propagated to infinity
using the multipolar potential of the cluster.

We have thus far discussed only elastic scattering. How can one deal
with inelastic collisions? Although this is unchartered territory, the
scenario will likely go as follows. In such a situation, new energy
channels open up. For each of these, there will be energy-diagonal
T-matrices $T_{d}(E_{i})$ and cross-energy ones
$T_{nd}(E_{i},E_{j})$. These will combine into a super-matrix
$T(E_{i},E_{j})=T_{d}(E_i)\delta_{E_{i},E_{j}}
+T_{nd}(E_{i},E_{j})$. The rest would basically be a repeat of what we
have done for a single channel. One would define the MS part
$\mathbf{T_c}$ with cut-offs on the range of the dipole and on the
angular momentum basis. Note that the latter is different for each
energy channel. Moreover, as the energy increases, larger and larger
angular momenta will be needed. Some adjustments on $a_c$ might also
be required in order to include inelastic channels involving more
extended molecular wavefunctions. Formally, Eq.~(\ref{eq:used}) still
applies to these super matrices. The resonant MS $(\mathbf{I}-\chi
\mathbf{T_c)}^{-1}$ term would not only exhibit peaks related to the
damped electronic modes of the cluster but also structures caused by
the inelastic contributions to $\mathbf{T_c}$ which will in turn
modulate the inelastic peaks in $\mathbf{T_{dip}}$. Interpolation of
the total CS would now involve linear interpolation of the type
described above in a multi-dimensional energy-channel space. The
computer time consuming aspect comes from repeated cluster
calculations of Eq.~(\ref{eq:used}) for many values of the incoming
electron energy $E_i$ and $l_{c}(E_i)$.
 
\ack

This research is supported by the Canadian Institutes of Health
Research and the EPSRC.\\

\appendix

\section{Relation between R and K matrices}

We suppose in the following, that the R-matrix defined at a radius
$r=a$ describes the inner region $r\leq a$ and that the potential
outside the sphere is zero. For elastic scattering we have
\cite{Burke1993},
\begin{equation}
F_{L}(a)=\sum_{L^{\prime }}R_{LL^{\prime }}\left[
r\frac{d}{dr}F_{L^{\prime }}\right] _{r=a}~.  \label{eq_R}
\end{equation}%
with $F_{L}$ the reduced radial wavefunctions.

For $r > a$, the wavefunction representing the scattering electron is%
\begin{equation}
\psi (\vec{r})=\sum_{L}\frac{F_{L}(r)}{r}Y_{L}(\hat{r})~.
\label{eq_psi}
\end{equation}

In the region $r>a$, one can write%
\begin{equation}
F_{L}(r)=f_{1L}~\bar{j}_{\ell }(\kappa r)+f_{2L}~\bar{\eta}_{\ell }(\kappa r)
\label{eq_F}
\end{equation}%
where $\bar{j}(z)=z~j(z)$ and $\bar{\eta}(z)=z~\eta (z)$. $F_{L}(r)$
is a solution of energy $E=\kappa ^{2}/2$ and angular momentum $L$ of
the Schr\"{o}dinger equation in a zero potential. Substituting
(\ref{eq_F}) in (\ref{eq_R}), we have%
\begin{equation}
f_{1L}~\bar{j}_{\ell }(\kappa a)+f_{2L}~\bar{\eta}_{\ell }(\kappa a)=a\kappa
\sum_{L^{\prime }}R_{LL^{\prime }}\left[ f_{1L^{\prime }}~\bar{j}_{\ell
^{\prime }}^{\prime }(\kappa a)+f_{2L^{\prime }}~\bar{\eta}_{\ell ^{\prime
}}^{\prime }(\kappa a)\right] ~,
\end{equation}%
where $\bar{j}_{\ell }^{\prime }(z)=d\bar{j}_{\ell }(z)/dz$ and $\bar{
\eta}_{\ell }^{\prime }(z)=d\bar{\eta}_{\ell }(z)/dz$. One can rearrange 
this equation and write it in matrix form%
\begin{equation}
\left(\mathbf{\bar{N}}-a\kappa \mathbf{R}\mathbf{\bar{N}^{\prime }}\right) f_{2}=-\left( \mathbf{\bar{J}}%
-a\kappa \mathbf{R} \mathbf{\bar{J}^{\prime }}\right) f_{1}
\end{equation}%
where $\bar{J}_{LL^{\prime }}=\bar{j}_{\ell }(\kappa a)\delta
_{LL^{\prime }}$, $\bar{N}_{LL^{\prime }}=\bar{\eta}_{\ell }(\kappa
a)\delta _{LL^{\prime }}$ and $f_{1},f_{2}$ are the amplitude column
vectors. This gives us:
\begin{equation}
f_{2}=-\left( \mathbf{\bar{N}}-a\kappa \mathbf{R} \mathbf{\bar{N}^{\prime }}\right) ^{-1}\left( \mathbf{\bar{J}}%
-a\kappa \mathbf{R} \mathbf{\bar{J}^{\prime }}\right) f_{1}  \label{eq_f}
\end{equation}%

We introduce the K-matrix, 
\begin{equation}
\mathbf{K}=\left( \mathbf{\bar{N}}-a\kappa \mathbf{R} \mathbf{\bar{N}^{\prime }}\right) ^{-1}\left( \mathbf{\bar{J}} -a\kappa \mathbf{R} \mathbf{\bar{J}^{\prime }}\right) ~.  \label{eq_K}
\end{equation}%

Substituting (\ref{eq_K}) in (\ref{eq_f}) and then in (\ref{eq_F}) and (\ref{eq_psi}%
), we have for $r>a$

\begin{equation}
\psi (\vec{r})=(\kappa r)^{-1}\sum_{LL^{\prime }}f_{1L}~\left[ \bar{j}_{\ell
}(\kappa r)\delta _{LL^{\prime }}-K_{L^{\prime }L}\bar{\eta}_{\ell ^{\prime
}}(\kappa r)\right] Y_{L^{\prime }}(\hat{r})
\end{equation}%
and thus%
\begin{equation}
\lim_{r\rightarrow \infty }\psi (\vec{r})=(\kappa r)^{-1}\sum_{LL^{\prime
}}f_{1L}~\left[ \sin (\kappa r-\ell \pi /2)\delta _{LL^{\prime
}}+K_{L^{\prime }L}\cos (\kappa r-\ell ^{\prime }\pi /2)\right] Y_{L^{\prime
}}(\hat{r})~,  \label{eq_psiK}
\end{equation}%
which is conform to the definition of the $\mathbf{K}$ matrix.

Eq.~(\ref{eq_K}) gives a 'passage' for the transformation $\mathbf{R}
\leftrightarrow \mathbf{K} $. It is worth noting that the R-matrix depends on
the radius but not the K-matrix (nor the transmission matrix $\mathbf{T}$, or the
scattering matrix $\mathbf{S}$). So once we have the R-matrix for any radius $a$
we can build the K-matrix using Eq.~(\ref{eq_K}) and inversely. The
T-matrix is then obtained from the K-matrix using the well known
relation:
\begin{equation}
\mathbf{T} = 2 i \mathbf{K} (1 - i \mathbf{K})^{-1}
\end{equation}%

\section{Surface to volume ratio}

One can estimate a surface to volume ration by the ratio of the number of 
surface molecules $N_{s}$ to the number of bulk molecules $N_{b}$. One can 
get an estimate of these numbers by considering a spherical cluster of radius 
$R_{c}$. The surface molecules are located in the shell between $R_{c}$ and 
$R_{c}-d$ where $d$ is, as above, the distance between molecules. If the mid-shell 
surface area $S=4 \pi (R_{c}-d/2)^{2}$ is occupied by molecules forming a close packed 
array, the area occupied by one molecule would be $S_{mol}\approx d^{2} \sqrt{3}%
/2$ and thus $N_{s}=S/S_{mol}$. Assuming the bulk of volume $V=4 \pi (R_{c}-d)^{3}/3$ 
is occupied by close packed molecules occupying a volume of $V_{b}=0.93 d^{3}%
\sqrt{3}/2$, one gets $N_{v}=V/V_{b}$. A high surface to volume ratio can thus 
be quantised by the quantity $SV=N_{s}/N_{b} \gg 1$. With $d=5.5$ a.u. one gets
$R_{c} \ll 4.6$. One thus deduces that the cluster must contain much less than
$N_{s}+N_{v} \approx 500$ molecules to have a large surface to volume ratio.

\newpage
\Tables

\begin{table}
{\centering\begin{tabular}{|c|c|c|c|}
\hline
{} & {H$_2$O}   & {(H$_2$O)$_2$}  \\
\hline
{r$_{OH1}$} & {1.81} & {1.810} \\
\hline
{r$_{OH2}$} & {1.81} & {1.823} \\
\hline
{$\theta_1$}    & {104.5} & {104.5}  \\
\hline
{r$_{OH3}$=r$_{OH4}$} & {}& {1.814} \\
\hline
{r$_{OO}$}  & {}     & {5.497} \\
\hline
{$\theta_2$}&{} & {104.6} \\
\hline
{$\alpha$} & {} & {4.7} \\
\hline
{$\beta$} & {}  & {55.1}  \\
\hline
\end{tabular}\small \par}
\caption{\label{tab:geometry} H$_2$O and  (H$_2$O)$_2$ geometries used in our 
calculations. The values of the dimer parameters (see figure~\ref{fig:geometry} for identification) are from \citeasnoun{PKK2001}. Length units are in bohr and angles in deg.}
\end{table}

\newpage
\Figures

\end{document}